\documentclass{article}

\PassOptionsToPackage{numbers, compress}{natbib}

\usepackage[final]{neurips_2021_ml4ps}

\usepackage[utf8]{inputenc} 
\usepackage[T1]{fontenc}    
\usepackage{hyperref}       
\usepackage{url}            
\usepackage{booktabs}       
\usepackage{amsfonts}       
\usepackage{nicefrac}       
\usepackage{microtype}      
\usepackage{xcolor}         

\usepackage{amsmath}
\usepackage{tabu} 
\usepackage{tikz}
\usepackage{tikzscale}
\usetikzlibrary{bayesnet}
\usepackage{subfig}
\usepackage{todonotes}
\usepackage{bm}

\title{CaloDVAE : Discrete Variational Autoencoders for Fast Calorimeter Shower Simulation}

\author{%
  Abhishek Abhishek$^1$\thanks{abhishek@myumanitoba.ca}, Eric Drechsler$^{1,2}$, Wojciech Fedorko$^1$, Bernd Stelzer$^{1,2}$ \\
  $^1$ TRIUMF, Vancouver, BC V6T 2A3 \\
  $^2$Simon Fraser University, Burnaby, BC V5A 1S6
}

\begin{document}

\maketitle

\begin{abstract}
    Calorimeter simulation is the most computationally expensive part of Monte Carlo generation of samples necessary for analysis of experimental data at the Large Hadron Collider (LHC). The High-Luminosity upgrade of the LHC would require an even larger amount of such samples. We present a technique based on Discrete Variational Autoencoders (DVAEs) to simulate particle showers in Electromagnetic Calorimeters. We discuss how this work paves the way towards exploration of quantum annealing processors as sampling devices for generation of simulated High Energy Physics datasets. 
\end{abstract}

\section{Introduction}
    \label{intro}
    With the advent of the High-Luminosity upgrade~\cite{calafiura2020atlas} and the expected increase in luminosity, experiments at the LHC are facing difficult computational challenges. A limiting factor on the precision of physics results is the lack of detailed Monte Carlo (MC) simulation in relevant phase spaces. This introduces a statistical uncertainty on measurements and hypothesis tests, limiting the sensitivity of LHC experiments. At the moment, billions of CPU hours~\cite{Karavakis:2014aia,Bozzi:1984010} are used by LHC experiments for MC simulation annually. 

The simulation of particles interacting with the calorimeter system is computationally demanding. In sampling calorimeters, particles interact electromagnetically or hadronically with dense absorber material, resulting in a cascade of subsequent particles - a particle shower. Active layers, like liquid Argon~\cite{McCarthy:2016pzc}, provide energy and location measurements from electric signals proportional to the number of particles produced in the cascade. The shower propagation and deposition of energy is an intrinsically probabilistic process. A full physics-based simulation 
with the state-of-the-art toolkit \textsc{Geant4}~\cite{GEANT4:2002zbu} can take minutes per event on current high-performance computing platforms~\cite{Aad_2010,Rahmat:2012fs}. Approximate algorithms based on parameterizations ~\cite{Grindhammer:1993kw,ATLAS:2010bfa} have been used in many applications and significantly decrease the runtime at the expense of accuracy. However, the developments of new methods involving hadronic and tau-jet sub-structure information may require the complete shower to be simulated~\cite{Dias:2016tea}, rendering such approximations insufficient.

Recent developments~\cite{de2017learning, paganini2018, ATL-SOFT-PUB-2018-001,atlascollaboration2021atlfast3} suggest, that deep generative models such as Generative Adversarial Networks (GANs) and Variational Autoencoders (VAEs) are able to provide approximations to underlying probability distributions of calorimeter shower data. Generating independent random samples from such models is computationally cheap, thus rendering them promising candidates for replacing parts of the default simulation framework. The ATLAS experiment has incorporated a GAN for calorimeter shower generation in a recent update to their simulation infrastructure~\cite{atlascollaboration2021atlfast3}.

Inspired by these remarkable successes, we introduce a Discrete Variational Autoencoder (DVAE) \cite{rolfe2016discrete,vahdat2018dvae++,khoshaman2018gumbolt} based model with hierarchical dependencies of latent variables in the approximate posterior and a Restricted Boltzmann Machine (RBM) latent prior. We study the qualitative performance of this model on an idealized calorimeter dataset \cite{calogan2017data} where electromagnetic calorimeter showers are simulated. We demonstrate that this model tackles the challenges brought on by the non-uniformly segmented nature of the calorimeter, dependence between energy deposits in sequential layers and varying sparsity of the activated calorimeter cells.

Quantum Variational Autoencoders (QVAEs) \cite{khoshaman2018quantum, winci2020path} are hybrid Quantum-Classical generative models which may be able to exploit quantum phenomena such as superposition and tunneling in quantum annealers to achieve better generative performance than their classical counterparts. This work is a first step towards applying QVAEs for calorimeter shower simulation and paves the way for future exploration and application of quantum annealing processors for generation of simulated High Energy Physics (HEP) datasets.



    
\section{Methodology}
    \label{methodology}
    \subsection{Dataset and Preprocessing}
\label{methodology:dataset}
We use the Electromagnetic Calorimeter Shower Images dataset \cite{calogan2017data} previously studied in \cite{paganini2018}. The dataset contains energy deposits from positrons, photons and charged pions in an idealised, longitudinally segmented EM calorimeter. An incident particle of certain type, energy and direction is generated and its interaction with the calorimeter material simulated using the \textsc{Geant4} 10.2.0 toolkit~\cite{GEANT4:2002zbu}  with the \textsc{Ftfp\_Bert} physics list ~\cite{Andersson:1996xi,Andersson:1986gw,Nilsson-Almqvist:1986ast,Ganhuyag:1997gz,Guthrie:1968ue,Bertini:1971xb,Karmanov:1979if} using the electromagnetic physics package~\cite{Burkhardt:2004ycp}. The calorimeter is a cube of volume $480$ mm$^3$  with three non-uniformly segmented layers. The exact geometry can be found in the appendix Table ~\ref{table:dimensions}.

In this work, we use a flattened representation where the energy deposits in each layer are "unrolled" and concatenated into a single feature vector for each example. The broad dynamic range of the energy deposited in a given calorimeter cell and the differences of energy deposit scales between different cells (e.g. cell near the middle vs near the edge of a layer) pose a challenge during training. Standardization is a common technique which makes data features approximately standard normally distributed, however is not suitable for highly sparse and either $0$ or strictly positive calorimeter shower data. We modify the standardization procedure to work with calorimeter shower data as described in Appendix \ref{appendix:dataset_standardization}.

\subsection{Deep Generative Models}

 \textbf{Variational Autoencoders (VAEs)}~\cite{kingma2014autoencoding, rezende2014stochastic} are a class of deep generative models that approximate the data distribution by optimizing an evidence lower bound (ELBO), 
 $\mathcal{L}_{\phi, \theta}(\mathbf{x})$ to the log-likelihood of the data under the model distribution, $\log{p_{\theta}(\mathbf{x})}$ : $$\mathcal{L}_{\phi, \theta}(\mathbf{x}) = \underbrace{\mathbb{E}_{q_{\phi}(\mathbf{z}|\mathbf{x})}[\log{p_{\theta}(\mathbf{x}}|\mathbf{z})]}_{\text{autoencoding term}} - \underbrace{\text{KL}[q_{\phi}(\mathbf{z}|\mathbf{x})||p(\mathbf{z})]}_{\text{kl term}} \leq \log{p_{\theta}(\mathbf{x})}$$
    
In the simplest case, the approximate posterior and prior over the latent variables are assumed to be factorized Gaussian distributions, $q_{\phi}(\mathbf{z}|\mathbf{x}) = \mathcal{N}(\mathbf{z}|\boldsymbol{\mu}_{\phi}(\mathbf{x}), diag(\boldsymbol{\sigma}^{2}_{\phi}(\mathbf{x}))$ and $p(\mathbf{z}) = \mathcal{N}(0, \mathbf{I})$ respectively. The approximate posterior, $q_{\phi}(\mathbf{z}|\mathbf{x})$ and generative, $p_{\theta}(\mathbf{x}|\mathbf{z})$ distributions are often parametrized using deep neural networks. The parameters $\phi$ and $\theta$ are optimized by minimizing the negative ELBO, $- \mathcal{L}_{\phi, \theta}(\mathbf{x})$ using stochastic gradient descent.

\textbf{Discrete Variational Autoencoders (DVAEs)} extend the VAE framework to allow discrete variables in the latent space ~\cite{rolfe2016discrete, vahdat2018dvae++, khoshaman2018gumbolt}. The non-differentiability of discrete variables does not allow for the reparameterization trick \cite{kingma2014autoencoding, rezende2014stochastic} to be used to compute low-variance gradient estimates of the autoencoding term w.r.t $\phi$. In this work, we focus on the GumBolt-DVAE model ~\cite{khoshaman2018gumbolt} which extends the Gumbel trick ~\cite{jang2016categorical, maddison2016concrete} for relaxing discrete distributions to work with Boltzmann machine (BM) priors. In GumBolt-DVAE, continuous proxy variables $\mathbf{\zeta}$ are used in replacement of discrete variables $\mathbf{z}$ during training while the discrete variables $\mathbf{z}$ are used during validation and generation.
The approximate posterior has a hierarchical structure, $q_{\phi}(\mathbf{z}|\mathbf{x}) = {\prod_{i}}q_{\phi_{i}}(\mathbf{z}_{i}|\mathbf{z}_{j < i}, \mathbf{x}), \mathbf{z} = [\mathbf{z}_{1}, \dots, \mathbf{z}_{N}]$ and the latent generative process is implemented by a restricted Boltzmann machine (RBM), $p_{\theta_{\textit{RBM}}}(\mathbf{z}) = e^{-E_{\theta_{\textit{RBM}}}(\mathbf{z})}/Z_{\theta} = e^{\mathbf{a}_\mathrm{l}^{T}\mathbf{z}_\mathrm{l} + \mathbf{a}_\mathrm{r}^{T}\mathbf{z}_\mathrm{r} +\mathbf{z}_\mathrm{l}^{T}\mathbf{W}\mathbf{z}_\mathrm{r}}/Z_{\theta}$, where $Z_{\theta}$ is the partition function. The RBM parameters ($a_{l}, a_{r}, W$) are jointly trained with the parameters ($\phi, \theta$). The complete set of latent variables predicted by the approximate posterior, $\mathbf{z} = \{\mathbf{z}_{1}, \dots, \mathbf{z}_{N}\}$ is partitioned into two equal subsets which form the two sides $\{\mathbf{z}_\mathrm{r}, \mathbf{z}_\mathrm{l}\}$ of the RBM. The hierarchical approximate posterior and RBM prior allow for rich latent space distributions and improve the generative performance of the model ~\cite{rolfe2016discrete, vahdat2018dvae++, khoshaman2018gumbolt}.
\subsection{CaloDVAE}

\begin{figure}[h!]
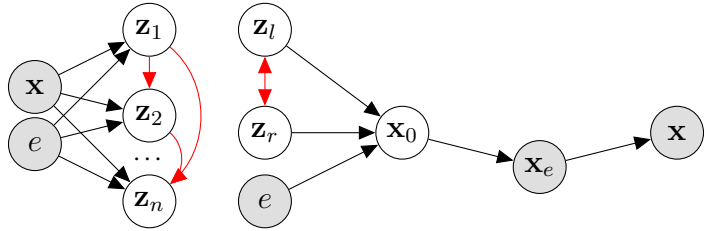

    \centering
    \subfloat{\includegraphics[height=3cm]{figures/inference_model}}
    \quad
    \subfloat{\includegraphics[height=3cm]{figures/generative_model}}
    \caption{Graphical description of the hierarchical approximate posterior $q_{\phi}(\mathbf{z}|\mathbf{x}, e)$ (left) and generative model $p_{\theta}(\mathbf{x}|\mathbf{z}, e)$ (right) to generate new synthetic samples. In the inference model (left), continuous proxies $\bm{\zeta}_{i}$ are used instead of discrete $\mathbf{z}_{i}$ during training. $e$ is the true or requested energy of the incident particle in GeV.}
    \label{fig:graphical_model}
\end{figure}

Our simulation technique is based on the GumBolt-DVAE framework. Figure \ref{fig:graphical_model} shows a graphical description of our model. We employ energy conditioning in a similar fashion to~\cite{ATL-SOFT-PUB-2018-001}. The approximate posterior distribution at a given hierarchy level $i$ and the generative distribution are specified as $q_{\phi_{i}}(\mathbf{z}_{i}|\mathbf{z}_{j < i}, \mathbf{x}, e)$ and $p_{\theta}(\mathbf{x}|\mathbf{z}, e)$ respectively, where $e$ is the true energy of the incident particle in GeV. Fully connected neural networks (FCNNs) with ReLU activation functions are used to parametrize $q_{\phi_{i}}(\mathbf{z}_{i}|\mathbf{z}_{j < i}, \mathbf{x}, e), i = 1, \dots, n$ and $p_{\theta}(\mathbf{x}|\mathbf{z}, e)$. In practice, since we use a flattened representation, the true incident particle energy, $e$ is simply concatenated to the input feature vector. The resulting vector, concatenated with $\{\mathbf{z}_{j < i}\}$ is passed through a sequence of non-linear fully connected layers to obtain approximate posterior samples $\mathbf{z}_{i}$ at a given hierarchy level $i$. During the autoencoding phase, approximate posterior samples $\{\mathbf{z}_{i}, i=1, \dots, n\}$ are concatenated with $e$ and passed through a sequence of non-linear fully connected layers to obtain a resampled version of the input $x$. During the generation phase to obtain new samples, RBM latent variable samples $\mathbf{z} \sim p_{\theta_{\textit{RBM}}}(\mathbf{z})$ obtained using block Gibbs sampling, concatenated with the requested incident particle energy $e$, are passed through a sequence of non-linear fully connected layers.

    

\paragraph{Output masking}
    Previous studies on the calorimeter dataset identified limitations in capturing the layer sparsity distributions in particular for charged pions\cite{paganini2018}. To overcome this, we introduce stochastic discrete variables $\mathbf{x}_{m}$ in the generative model where each $x_{m, i} \in \{0, 1\}$ determines whether the calorimeter cell $i$ is hit. These variables, in addition to the generated energy deposits $\mathbf{x}_{e}$ are used to produce the final output, $\mathbf{x} = \mathbf{x}_{m} \odot \mathbf{x}_{e}$, $\odot$ denotes the Hadamard product. In practice, a hidden vector $\mathbf{x}_0$ is first obtained by passing either approximate posterior samples or RBM samples concatenated with the incident particle energy through a sequence of non-linear fully connected layers. $\mathbf{x}_0$ is then passed through a second set of non-linear fully connected layers to obtain $\mathbf{x}_{m}$ and $\mathbf{x}_{e}$ independently. A ReLU activation function is applied to $\mathbf{x}_{e}$ since $\forall i, x_{e_{i}} \geq 0$ and to encourage sparsity ~\cite{de2017learning, paganini2018}. We use the Gumbel trick ~\cite{jang2016categorical, maddison2016concrete} for discrete variables $\mathbf{x}_{m}$ during training to ensure differentiability (i.e. continuous proxy variables are used instead during training). Binary Cross Entropy (BCE) loss applied to $\mathbf{x}_{m}$ and Mean Squared Error (MSE) loss applied to $\mathbf{x}$ are summed to compute the total autoencoding loss used to train the model.

\section{Preliminary Results}
    \label{results}
    We performed a grid search to heuristically determine the best hyperparameter setting for each particle type. A separate model with the best hyperparameter setting is trained for each particle type  and used to produce the results. We include the details on the hyperparameter scans and settings used to produce the following results in Appendix \ref{appendix:hyperparameter_scan}.

\begin{figure}[h!]
    \centering
    \subfloat{\includegraphics[width=0.25\linewidth]{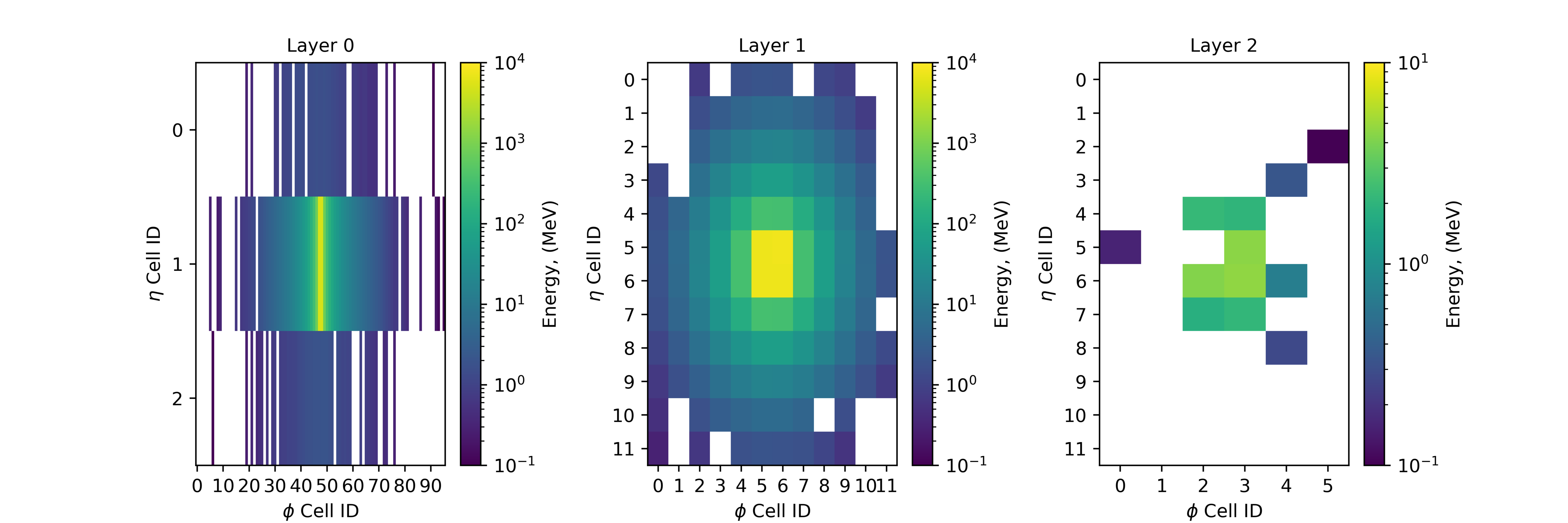}}
    \subfloat{\includegraphics[width=0.25\linewidth]{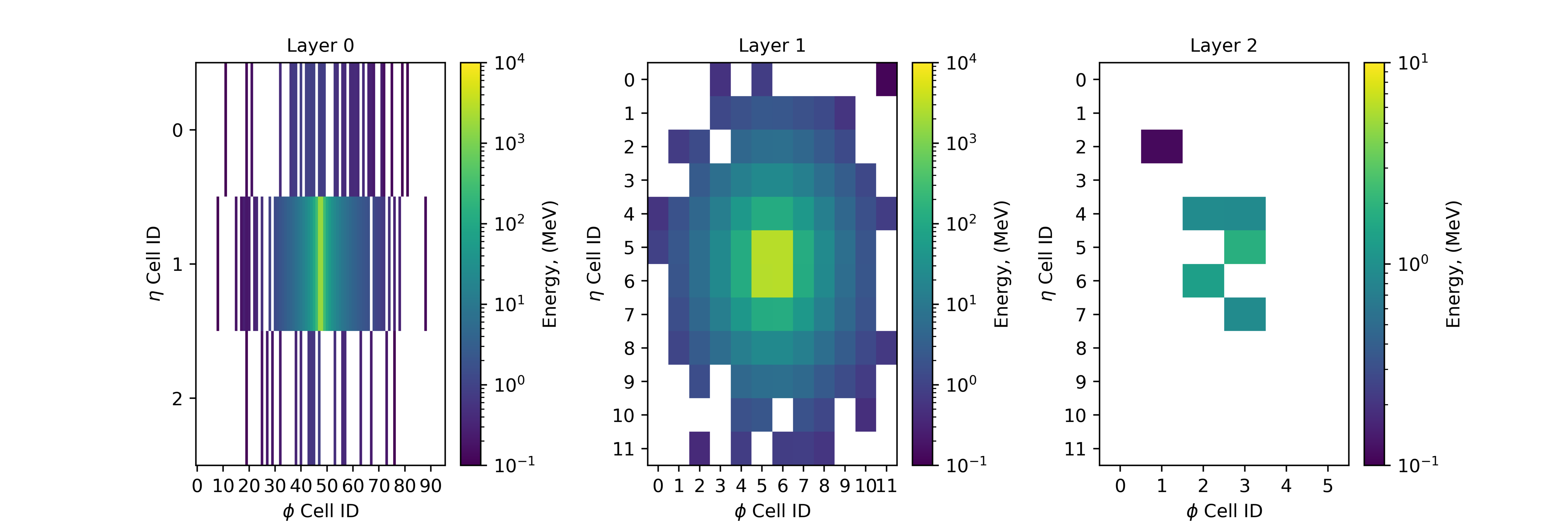}}
    \subfloat{\includegraphics[width=0.25\linewidth]{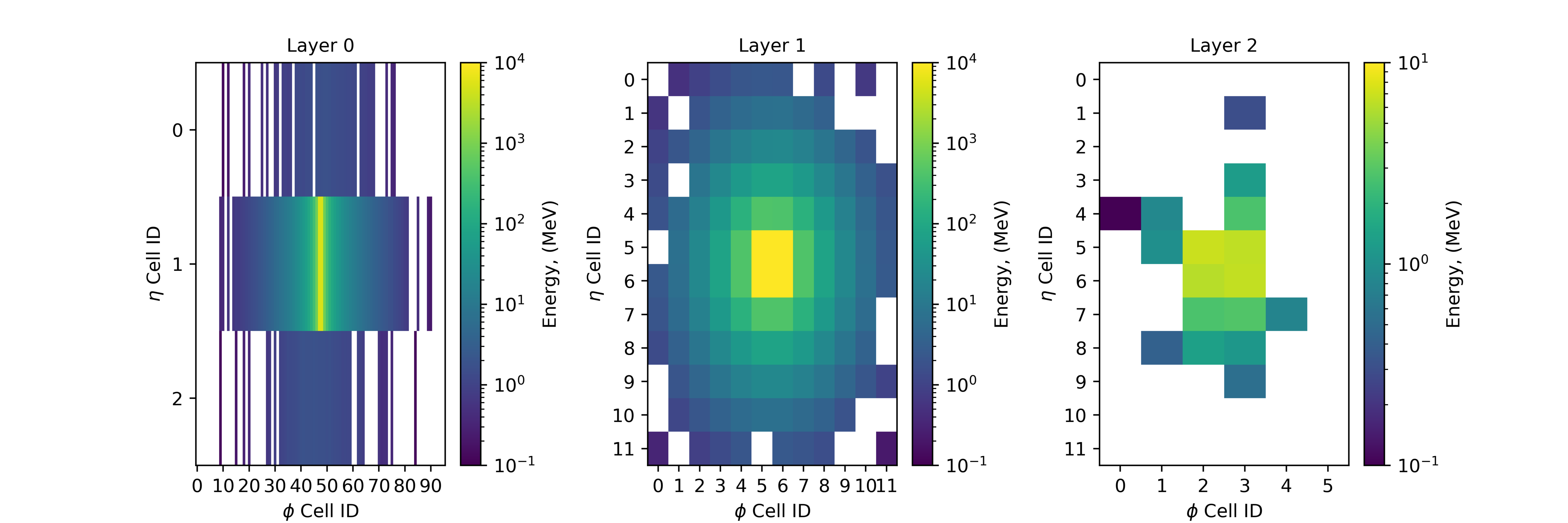}}
    \subfloat{\includegraphics[width=0.25\linewidth]{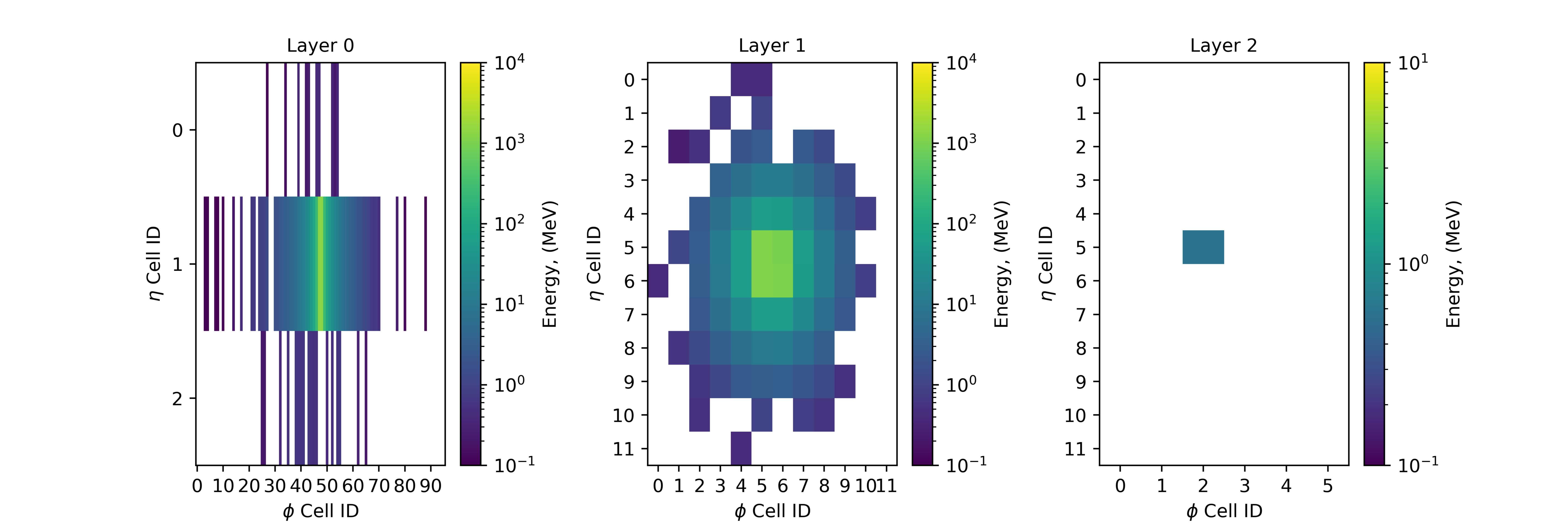}}
    \newline
    \subfloat{\includegraphics[width=0.25\linewidth]{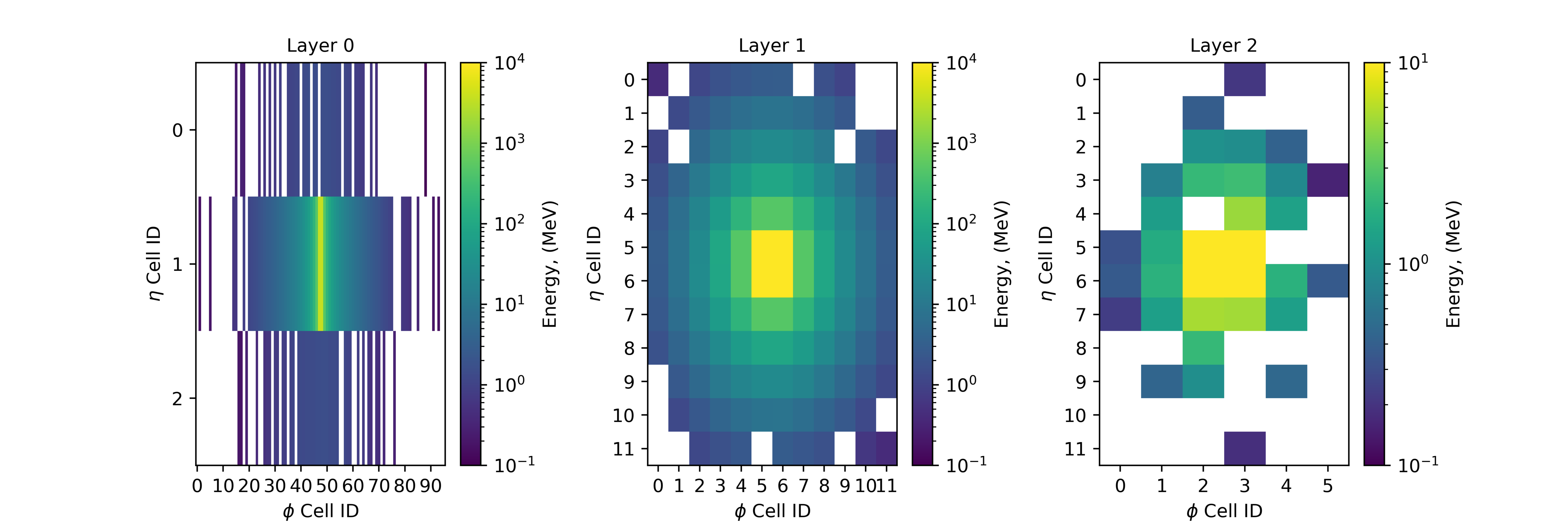}}
    \subfloat{\includegraphics[width=0.25\linewidth]{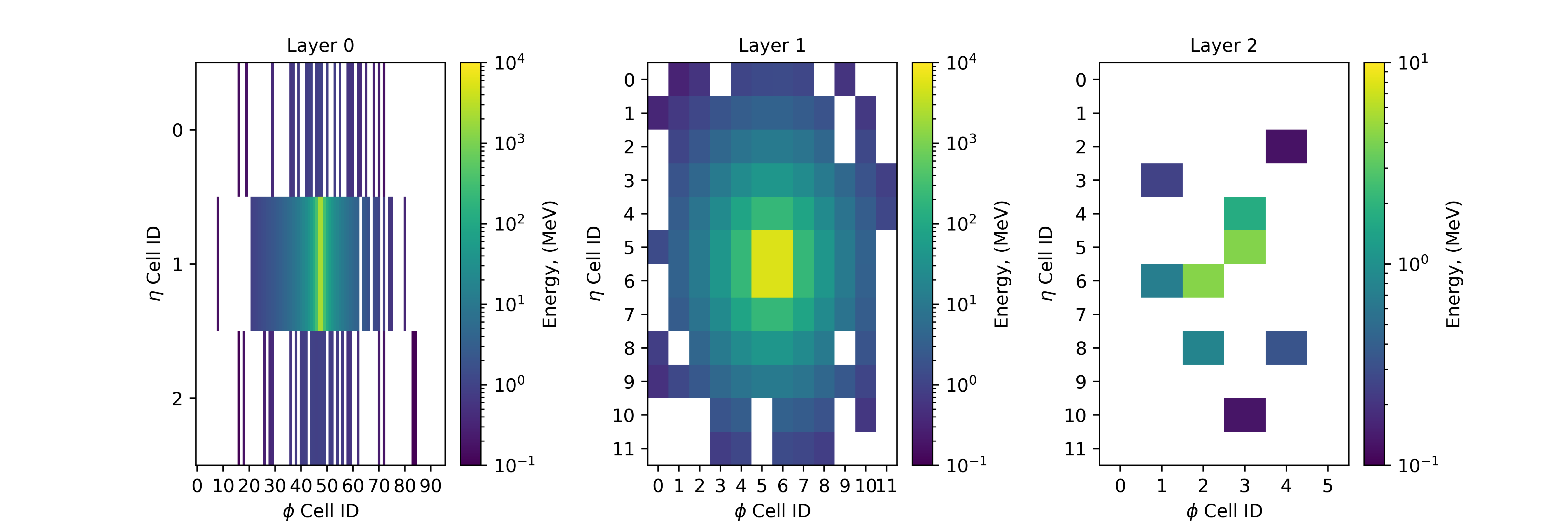}}
    \subfloat{\includegraphics[width=0.25\linewidth]{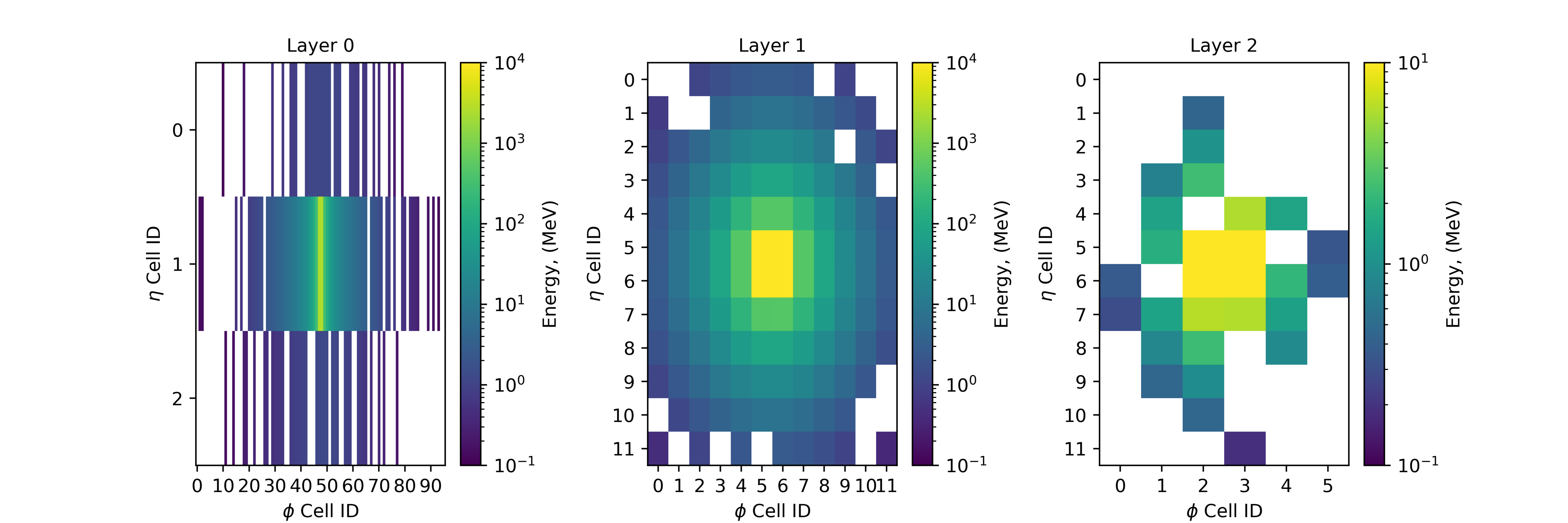}}
    \subfloat{\includegraphics[width=0.25\linewidth]{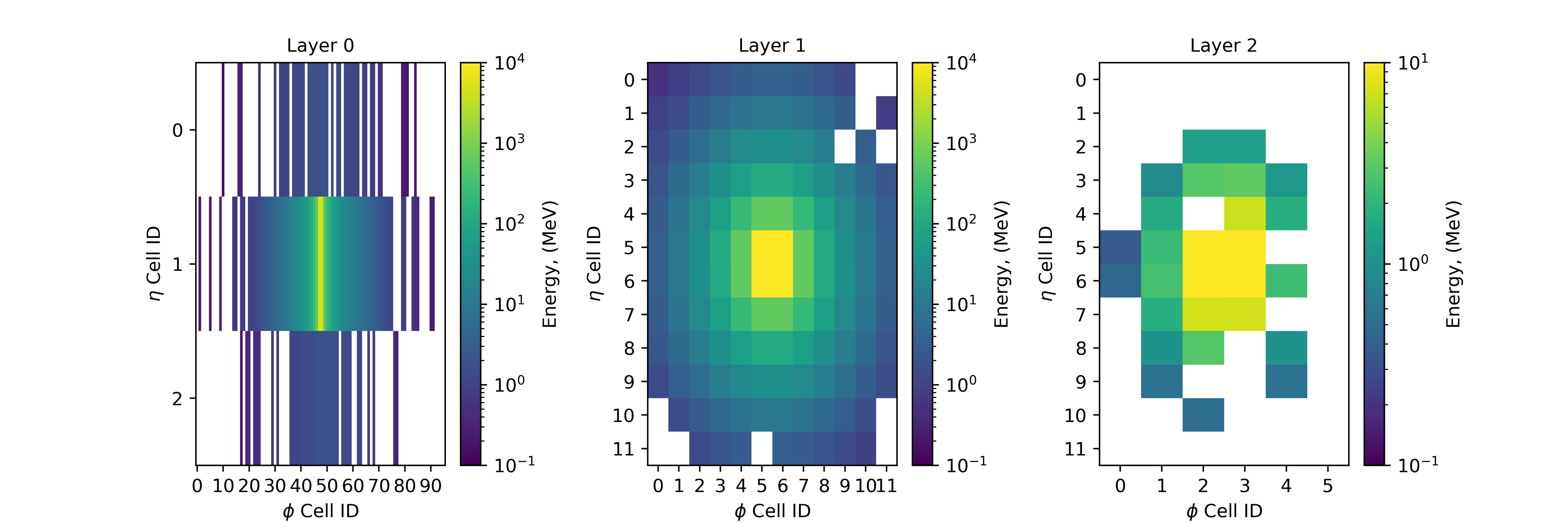}}
    \newline
    \subfloat{\includegraphics[width=0.25\linewidth]{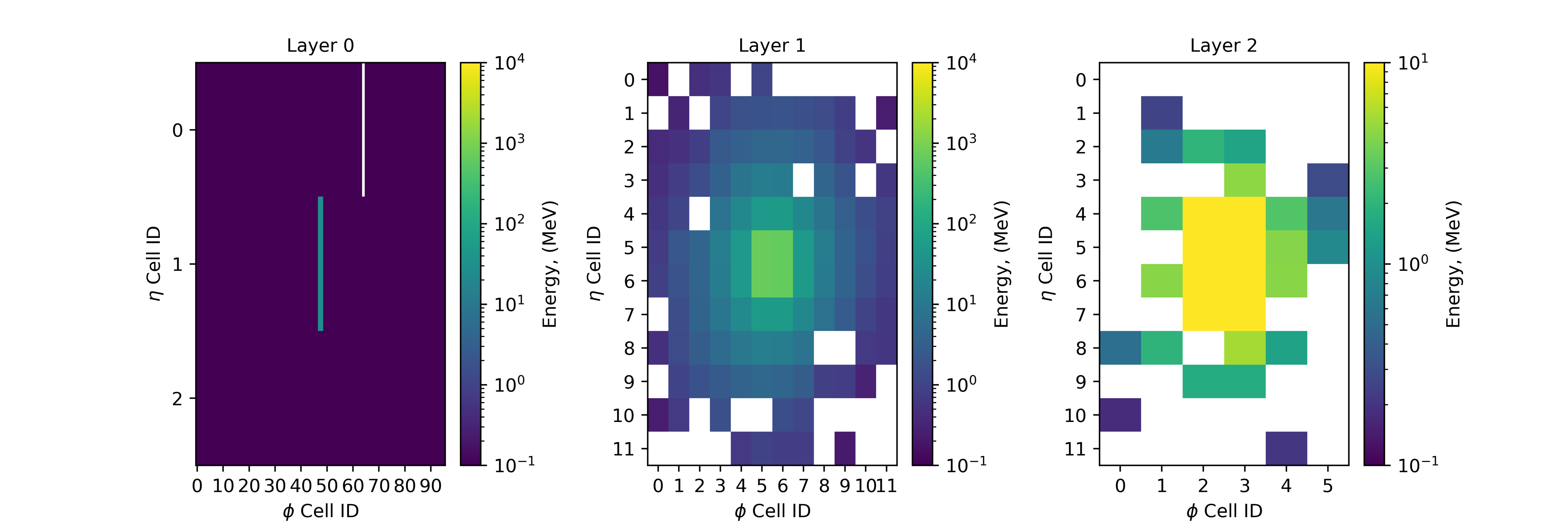}}
    \subfloat{\includegraphics[width=0.25\linewidth]{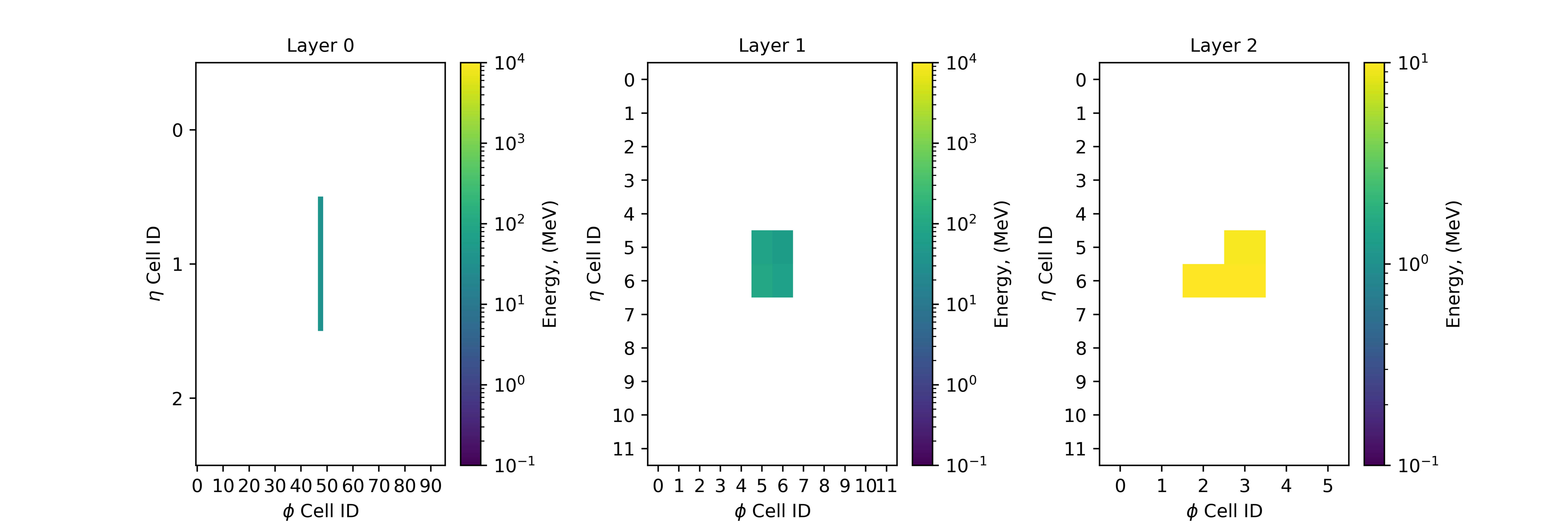}}
    \subfloat{\includegraphics[width=0.25\linewidth]{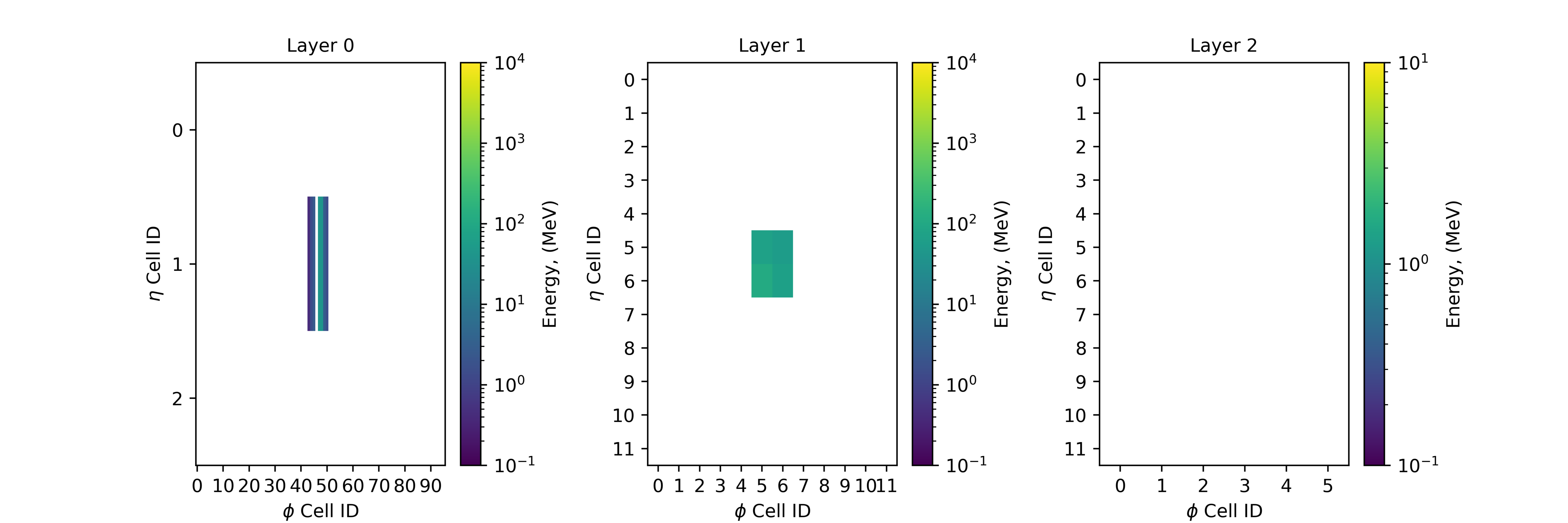}}
    \subfloat{\includegraphics[width=0.25\linewidth]{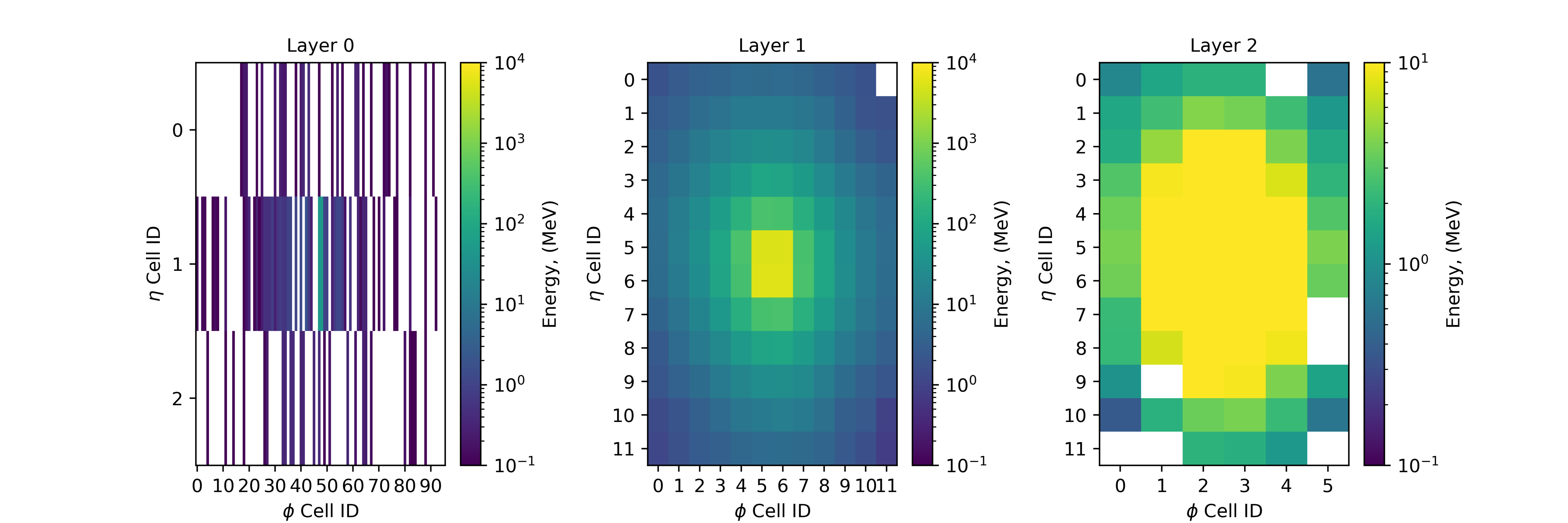}}
    \caption{Examples of shower images of CaloDVAE samples for $e^{+}$ (top), $\gamma$ (middle) and $\pi^{+}$ (bottom) with incident particle energy $e \sim \mathcal{U}[0, 100]$ GeV.}
    \label{fig:shower_images}
\end{figure}

\textbf{Qualitative assessment} of shower images of CaloDVAE samples (Appendix \ref{appendix:shower_images}, Figure \ref{fig:shower_images}) reveals that a broad variety of samples are generated by our model, reproducing features such as the patterns of activated and non-activated cells, centrality and lateral width of the clusters, as well as longitudinal behaviour of the shower. Of note is the behaviour in the last two layers in the pion sample, where some of the generated showers have very little deposit in these two layers, whereas in other samples large energy deposit is seen - a feature observed in the simulated training data. Our model also displays good energy conditioning and extrapolation behavior beyond the energy region in which it has been trained (\textit{cf.} Appendix~\ref{appendix:energy_conditioning}).

\textbf{Shower shape variables} are determined by the transverse and longitudinal profile of the shower, and are useful for particle identification and energy calibration ~\cite{paganini2018}. We present 1D histograms for a subset of these variables in Figure \ref{fig:shower_shape_variable_histograms} and their description in Appendix \ref{appendix:shower_shape_variables}, Table \ref{table:qualityvariables}. GEANT4 samples from the test subset of the dataset and CaloDVAE samples with $e \sim \mathcal{U}[1, 100]$ GeV were used to fill the histograms. The shower shape distributions approximately match at different scales and in particular, layer sparsity distributions which were previously observed to be challenging are recovered faithfully. Correct modelling of the bi-modal sparsity distribution for charged pions in layer 1 is quite notable since they undergo both hadronic and electromagnetic interactions.
    
\begin{figure}[ht!]
    \centering
    \includegraphics[width=\textwidth]{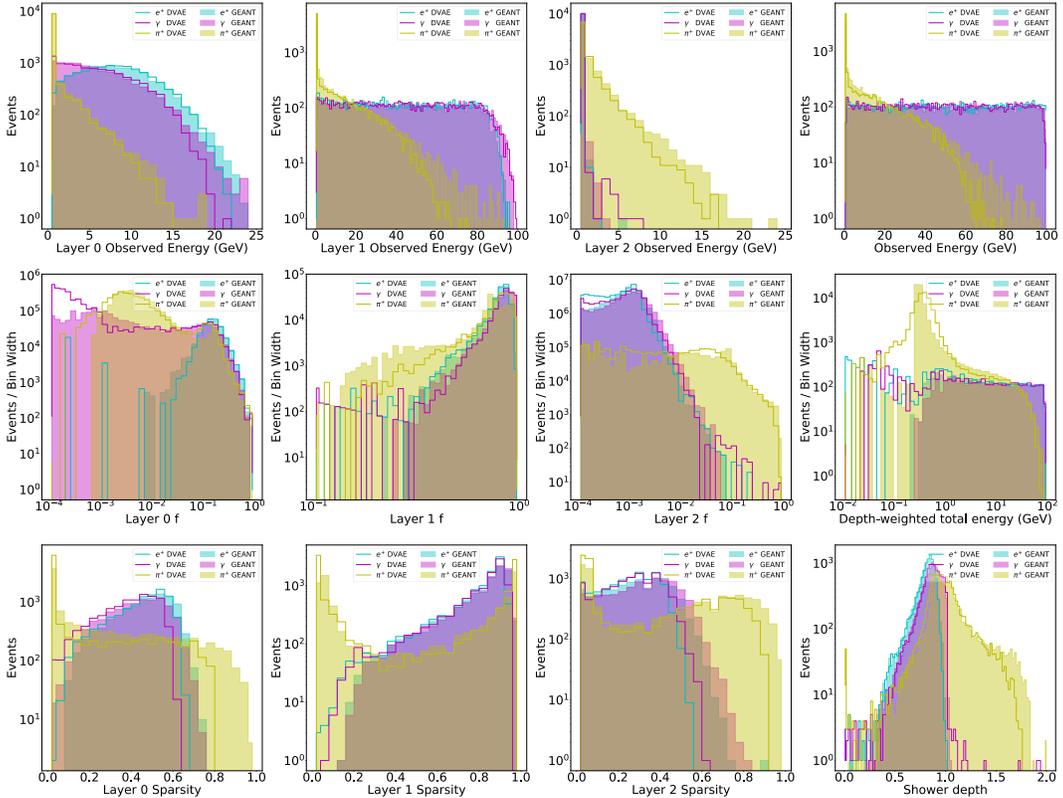}
    \caption{Shower shape variables (Appendix \ref{appendix:shower_shape_variables}, Table \ref{table:qualityvariables}) and layer sparsity (fraction of cells hit) distributions for GEANT4 and CaloDVAE samples.}
    \label{fig:shower_shape_variable_histograms}
\end{figure}
    
\section{Discussion and Future Outlook}
    \label{discussion}
    \paragraph{QVAE}
\label{discussion:qvae}
In QVAEs, a Quantum Boltzmann machine (QBM) \cite{amin2018quantum} replaces the restricted Boltzmann machine (RBM) implementing the latent generative process in the DVAE framework ~\cite{khoshaman2018quantum, winci2020path} and offloads the classical latent space sampling to a quantum annealer. Quantum annealers operated as sampling devices may provide a computational advantage over Markov Chain Monte Carlo (MCMC) techniques when using large latent-space BMs. Previous work ~\cite{winci2020path} has shown remarkable success on the MNIST and FMNIST datasets but notes that more complex datasets are required to fully exploit the large BMs in the latent space. The successful reproduction of high level physics observables by our model indicates that models of this class have high enough expressibility (enabled by the trainable complex prior) to model High Energy Physics (HEP) datasets such as calorimeter showers. Therefore, generative modelling of HEP datasets may benefit from using quantum annealers as Boltzmann sampling devices. Our work provides a template for application of similar techniques to other HEP datasets and is a necessary first step towards the exploration and application of quantum annealing processors for generation of simulated HEP datasets.
    
\section{Broader Impact}
    \label{impact}
    We considered potential negative impacts of the research presented. Indirectly, the method presented can be used to create deceptive fake data - a concern common to most generative methods. We note that this concern already exists with the foundational works~\cite{rolfe2016discrete, vahdat2018dvae++, khoshaman2018gumbolt} upon which this application work builds. Within the presented application domain the negative impact of the work may include production of biased simulation samples and thus affecting scientific results that rely on these samples. This concern would also apply to any traditional or novel method of generating simulated data. 

We believe the potential negative impacts are offset by the positive impacts on science and society. If the full potential of the work presented here is eventually realized - i.e. if quantum processors can be harnessed for the generation of synthetic data - millions of CPU years per year could be saved that would have to be otherwise devoted to the task of simulated data generation for the HL-LHC experiments. This saving can have enabling impact in terms of fiscal considerations, but also will contribute to reduced environmental footprint. The sensitivity of physics analysis could also be improved through the availability of large synthetic datasets. We also note the potential for the development of semi-supervised methods based on the methodology presented, thus enabling learning on real experimental data and potential reduction of systematic uncertainties in the final physics analyses in HEP experiments.
    
\section*{Acknowledgements}
    \label{acknowledge}
    We gratefully acknowledge the support of NSERC and Compute Canada. The authors would like to thank Olivia Di Matteo, Mohammad Amin and Walter Vinci for helpful discussions. Weights and Biases with an academic license was used for experiment tracking ~\cite{wandb}.

\bibliography{references}

\section*{Checklist}
    \begin{enumerate}

\item For all authors...
\begin{enumerate}
  \item Do the main claims made in the abstract and introduction accurately reflect the paper's contributions and scope?
    \answerYes{We believe the scope of the work is accurately represented in the abstract.}
  \item Did you describe the limitations of your work?
    \answerYes{See section~\ref{discussion}}
  \item Did you discuss any potential negative societal impacts of your work?
    \answerYes{See section~\ref{impact}}
  \item Have you read the ethics review guidelines and ensured that your paper conforms to them?
    \answerYes{We submit that the work presented adheres to the ethical standards outlined in the 'Ethics Guidelines' document}
\end{enumerate}

\item If you are including theoretical results...
\begin{enumerate}
  \item Did you state the full set of assumptions of all theoretical results?
    \answerNA{}
	\item Did you include complete proofs of all theoretical results?
    \answerNA{}
\end{enumerate}

\item If you ran experiments...
\begin{enumerate}
  \item Did you include the code, data, and instructions needed to reproduce the main experimental results (either in the supplemental material or as a URL)?
    \answerNo{Release of the code at this time proved impractical as the work is progressing toward more advanced applications}
  \item Did you specify all the training details (e.g., data splits, hyperparameters, how they were chosen)?
    \answerYes{We outline the method of data pre-processing and hyperparameters used and methodology for selecting the final choice in Appendices~\ref{appendix:dataset_standardization} and~\ref{appendix:hyperparameter_scan}}
	\item Did you report error bars (e.g., with respect to the random seed after running experiments multiple times)?
    \answerNo{We do not repeat the training sessions with multiple random seeds as precise quantification of the agreement of the generated dataset with the primary samples is beyond the scope of the paper. We argue that quantification of such agreement over multi-dimensional feature space is challenging in principle, and generally not done in similar expense. Even barring this the toy nature of the dataset would not justify the expense of repeat trials to estimate the 'error bar' on such agreement. The scope of the paper includes only the assesment if the discrete latent variable generative models are capapble in principle of reproducing complex physics distributions.}
	\item Did you include the total amount of compute and the type of resources used (e.g., type of GPUs, internal cluster, or cloud provider)?
    \answerNo{We did not rigorously track resource and computing time used on the variety of assets we used for training and evaluation of our model as this is out of scope of the current state of the study - which concentrated on establishing the feasibility of discrete latent space model to generate complex physics datasets. We do not claim that processing/sample generation time of this particular model would be competitive in relation to other generative models. This is because of the Monte Carlo sampling stage inherently built into sampling of the RBM - however the application of quantum processors is expected to address this.}
\end{enumerate}

\item If you are using existing assets (e.g., code, data, models) or curating/releasing new assets...
\begin{enumerate}
  \item If your work uses existing assets, did you cite the creators?
    \answerYes{See Section \ref{methodology:dataset}}
  \item Did you mention the license of the assets?
    \answerNo{We do not mention the license, however it is easily found by following the reference in the text.}
  \item Did you include any new assets either in the supplemental material or as a URL?
    \answerNo{}
  \item Did you discuss whether and how consent was obtained from people whose data you're using/curating?
    \answerNA{Data is available under CC BY 4.0 licence}
  \item Did you discuss whether the data you are using/curating contains personally identifiable information or offensive content?
    \answerNA{}
\end{enumerate}

\item If you used crowdsourcing or conducted research with human subjects...
\begin{enumerate}
  \item Did you include the full text of instructions given to participants and screenshots, if applicable?
    \answerNA{}
  \item Did you describe any potential participant risks, with links to Institutional Review Board (IRB) approvals, if applicable?
    \answerNA{}
  \item Did you include the estimated hourly wage paid to participants and the total amount spent on participant compensation?
    \answerNA{}
\end{enumerate}

\end{enumerate}

\section{Appendix}
    \label{appendix}
    \subsection{Dataset standardization}
\label{appendix:dataset_standardization}

Applying standardization to data features makes them standard normally distributed by removing the mean and scaling to unit variance. However, in the case of calorimeter shower data,  a value of $0$ for a given cell denotes a non-hit cell. Since the raw values in our data correspond to energy depositions, they are either $0$ or strictly positive. To achieve some form of standardization while preserving the "physical" characteristics, we scale each cell independently using the following steps.

For a given cell $i$, let $X_{i\neq 0}$ and $X_{i=0}$ denote the sets of dataset samples in which cell $i$ is hit and not hit respectively. The values of cell $i$ for samples in $X_{i\neq 0}$ are standardized using mean $\mu_{i}$ and variance $\sigma_{i}^{2}$ computed over $X_{i\neq 0}$. Additionally, to maintain a distinction b/w samples in which cell $i$ is hit and not hit, we shift the values of cell $i$ for samples in $X_{i\neq 0}$ by their smallest value plus an $\epsilon$ if the smallest value is negative. $\epsilon$ is a small positive number, e.g. 0.01. This allows to maintain a distinction between hit and non-hit cells.

\subsection{Shower images}
An example in this dataset can be represented as 3 grayscale 2D images in the ($\eta-\phi$) space, where $\eta$ is the beam direction in an experiment and $\phi$ is direction perpendicular to both $\eta$ and $z$, the particle propagation direction. The intensity of a pixel is the amount of energy deposited in the corresponding calorimeter cell.

\label{appendix:shower_images}

\subsection{Shower Shape Variables}
\label{appendix:shower_shape_variables}
\begin{table*}[h]
    \centering
    \renewcommand{\arraystretch}{1.2}
    \begin{tabular}{ m{4.2cm}  m{5.1cm}  m{4.3cm}} 
        \toprule
         {Shower Shape Variable} & {Formula} & \textit{Notes}
         \\
         \midrule
         $E_i$ & $E_i = \sum_\mathrm{pixels} \mathcal{I}_i $ & Energy deposited in the $i^{th}$ layer of calorimeter\\
         \midrule
         $E_\mathrm{tot}$ & $E_\mathrm{tot}=\sum\limits_{i=0}^2 E_i$ & Total energy deposited in the electromagnetic calorimeter\\
         \midrule
         $f_i$ & $f_i = E_i / E_\mathrm{tot}$ & Fraction of measured energy deposited in the $i^{th}$ layer of calorimeter\\
        \midrule
        Depth-weighted total \newline energy, $l_d$ & $l_d=\sum\limits_{i=0}^2 i\cdot  E_i$ & The sum of the energy per layer, weighted by layer number \\
        \midrule
        Shower Depth, $s_d$ & $s_d = l_d / E_\mathrm{tot}$ & The energy-weighted depth in units of layer number \\
        \bottomrule
    \end{tabular}
    \caption{Variables characterizing the properties of the simulated showers. \cite{paganini2018}}
    \label{table:qualityvariables}
\end{table*}
\cleardoublepage

\subsection{Hyperparameter Scan}
\label{appendix:hyperparameter_scan}

We split the 100,000 GEANT4-simulated event dataset using a 80\%-10\%-10\% split for train, validation, and test subsets, respectively. An optimal setting for the model and training parameters was determined heuristically independently for each incident particle type. Three qualitatively different model settings were investigated; their specifications are listed in Table~\ref{table:models}. The models' performances were evaluated over a grid of training hyperparameters, summarised in Table~\ref{table:hyppargrid}. In order to determine the best parameter sets, the distributions of shower shape variables for GEANT4 samples from the validation subset of the dataset and CaloDVAE samples were compared using a Kolmogorov-Smirnov (KS) test. The optimal setting selected as the one maximising the KS probability over the complete set of shower variables.

It was observed that for all particle types the model architecture III and IV generalised best. This indicates that additional depth in encoder and decoder, as well as an increase in dimensionality in the latent space provides a more powerful model, capturing the underlying dataset complexity more efficiently.

 \begin{table*}[ht!]
 \centering
\renewcommand{\arraystretch}{1.2}
\begin{tabular}{ l l l m{2cm} m{1.5cm}} 
\toprule
{Parameter} & Encoder Layers  & Decoder Layers & Hierarchy Levels  & Latent Nodes Per \\
 &  & & & Hierarchy Level \\
\midrule
Model I & $[400,300,200]$ & $[200,300,400]$ & $2$ & $64$ \\
Model II & $[400,350,300,200]$ & $[200,300,350,400]$ & $4$ & $128$ \\
Model III & $[400,350,300,250,200]$ & $[200,250,300,350,400]$ & $4$ & $128$ \\
Model IV & $[500,450,400,350,300]$ & $[300,350,400,450,500]$ & $6$ & $150$ \\
\bottomrule
\end{tabular}
\caption{Different model architectures explored in the hyper-parameter scan. Each successive model grows in complexity by adding layers in encoder and decoder, introducing additional hierarchy levels and increasing the latent space dimensionality.}
\label{table:models}
\end{table*}

\begin{table*}[ht!]
\centering
\renewcommand{\arraystretch}{1.1}
\begin{tabular}{cc} 
\toprule
 {Parameter} & {Range} \\
\midrule
Learning Rate & $[0.01,0.005,1.e^{-3},10^{-4},0.5\times 10^{-4}]$ \\
Epochs & $[25,50,75,100]$ \\
Batch Size & $[50,64,75,100,128,192]$ \\
Latent smoothing temp. $\tau_{\mathbf{z}}$   & $[1/5,1/7,1/9]$ \\
Output mask smoothing temp. $\tau_{\mathbf{x}_{m}}$ &  $[1/5,1/7,1/9]$  \\
\bottomrule
\end{tabular}
\caption{Grid of parameters considered for the hyperparameter optimization using the three model definitions in Table~\ref{table:models}. Latent smoothing temp. $\tau_{\mathbf{z}}$ and output mask smoothing temp. $\tau_{\mathbf{x}_{m}}$ are parameters of the Gumbel trick used to control the "smoothness" of the continuous proxy variables. ~\cite{jang2016categorical, maddison2016concrete}}
\label{table:hyppargrid}
\end{table*}

\begin{table*}
\centering

\renewcommand{\arraystretch}{1.1}
\begin{tabular}{lccc} 
\toprule
 & Positron $e^+$ & Photon $\gamma$ & Charged Pion $\pi^+$ \\
\midrule
Model Type & Model II &  Model IV &  Model IV  \\
Learning Rate & $10^{-4}$ & $0.5\times10^{-4}$ & $10^{-4}$\\
Epochs & $100$ & $100$ & $100$ \\
Batch Size & $100$& $100$& $100$ \\
Latent smoothing temp. $\tau_{\mathbf{z}}$  & $1/5$ & $1/7$ & $1/5$ \\
Output mask smoothing temp. $\tau_{\mathbf{x}_{m}}$ & $1/5$ & $1/5$ & $1/9$ \\
\bottomrule
\end{tabular}
\caption{Selected Models per incident particle type after heuristic evaluation of the hyperparameter scan used to produce the preliminary results. For definitions and ranges, see Table~\ref{table:models} and Table~\ref{table:hyppargrid}.}
\label{table:selectedmodels}
\end{table*}

\begin{table}[ht!]
\centering
\renewcommand{\arraystretch}{1.1}
\begin{tabular}{lccc} 
 \toprule
  Layer  & $\Delta z$ [mm]  & $\Delta \eta$ [mm] & $\Delta \phi$ [mm] \\ 
 \midrule
 0 & 90 & 5 & 160 \\ 
 1 & 347 & 40 & 40 \\
 2 & 43 & 80 & 40 \\
 \bottomrule
\end{tabular}
\caption{The geometry of the calorimeter. The $z$-axis corresponds to the direction of particle propagation, the $\eta$- and $\phi$-axes are perpendicular to this \cite{paganini2018}.}
\label{table:dimensions}
\end{table}


\subsection{Energy conditioning}
\label{appendix:energy_conditioning}
\begin{figure}[h!]
    \centering
    \includegraphics[width=0.5\textwidth]{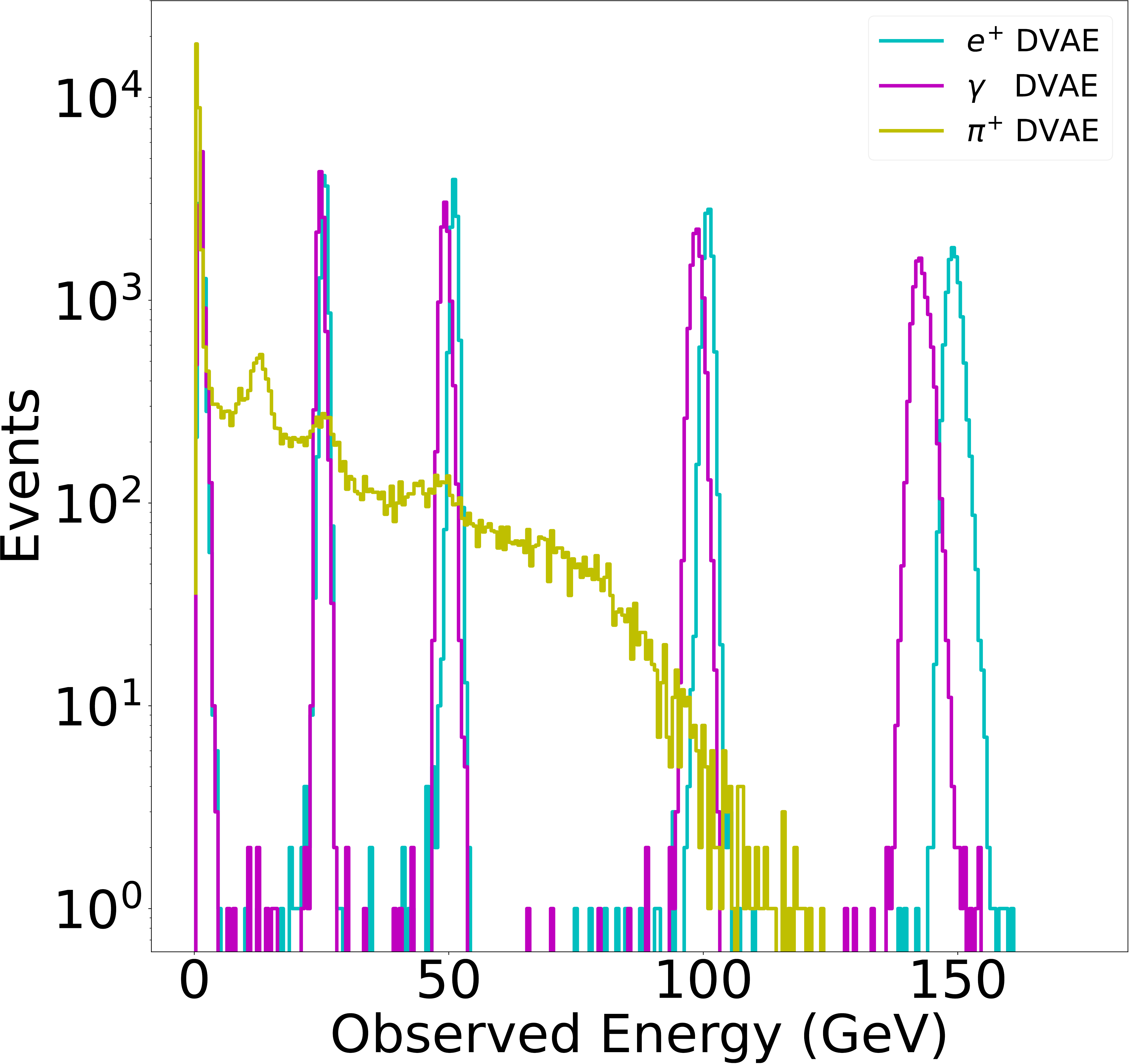}
    \caption{Observed energy spectra for synthetic CaloDVAE $e^{+}, \gamma$ and $\pi^{+}$ samples generated with true incident energies of 1, 25, 50, 100 and 150 GeV.}
    \label{fig:sample_hist}
\end{figure}

We study the performance of energy conditioning of the model. We sample from the model requesting specific values of true energy (1, 25, 50, 100 and 150 GeV) of the incident particle and histogram total observed energy in the cluster. As shown in Fig~\ref{fig:sample_hist} photon and electron clusters display sharp peaks at the requested energy values, whereas pion samples display broadened response - however this is due to the nature of the uncontained charged pion shower in the electromagnetic calorimeter and not due to poor model conditioning. Note that the 150 GeV exceedes the energy range where the models have been trained.
\end{document}